\documentclass[preprint,sort&compress,11pt]{elsarticle}
\journal{Journal Of Computational Physics}

\usepackage{latexsym}
\usepackage{amsmath}
\usepackage{graphicx,color}
\usepackage{hyperref}
\usepackage{alltt,bm}
\usepackage{verbatim}
\newcommand{\nod}{\ensuremath{\mathcal{N}}}
\newcommand{\pr}{\ensuremath{\mathcal{P}}}
\newcommand{\ix}{\ensuremath{I_{l\rightarrow g}}}
\newcommand{\ixx}{\ensuremath{I^p_{l\rightarrow g}}}
\newcommand{\cc}{\ensuremath{c}}
\usepackage[ruled,vlined,commentsnumbered]{algorithm2e}

\usepackage[normalem]{ulem}

\begin{document}

\begin{frontmatter}

\title{GPU accelerated spectral finite elements on all-hex meshes}

\author[a,b]{J.-F. Remacle}
\author[a]{R. Gandham}
\author[a]{T. Warburton}
\address[a]{Department of Computational and Applied Mathematics, Rice University}
\address[b]{Universit\'e catholique de Louvain, Institute of Mechanics,
Materials and Civil Engineering (iMMC), B\^atiment Euler,
Avenue Georges Lema\^itre 4, 1348 Louvain-la-Neuve, Belgium}

\begin{abstract}
This paper presents a spectral element finite element scheme that
efficiently solves elliptic problems on unstructured hexahedral
meshes. The discrete equations are solved using a matrix-free
preconditioned conjugate gradient algorithm. An additive Schwartz two-scale preconditioner
is employed that allows h-independence convergence. An extensible
multi-threading programming API is used as a common kernel language
that allows runtime selecti	on of different computing devices (GPU and CPU)
and different threading interfaces (CUDA, OpenCL and OpenMP).
Performance tests demonstrate that problems with over 50 million
degrees of freedom can be solved in a few seconds on an off-the-shelf GPU.
\end{abstract}

\end{frontmatter}

\pagestyle{myheadings}
\thispagestyle{plain}

\section{Introduction}
Recent research efforts \cite{yamakawa2003fully,baudouin2014frontal} have led to
the development of 3D hex-dominant mesh generation systems that are
fast and reliable. It is now possible (e.g. with Gmsh \cite{geuzaine2009gmsh}) to
create meshes of general 3D domains that contain over 80 \% of
hexahedra in volume in a fully automatic manner.

We foresee that fully automatic hex-meshing
procedures will be available in the next decade. This perspective
allows finite element researchers to reconsider some commonly held beliefs, namely that
tet-meshing may not remain the only solution for automatic
mesh generation.

Quadrilateral meshes in 2D and hexahedral meshes in 3D are usually
considered to be superior to triangular/tetrahedral meshes.
There are numerous modeling reasons to prefer hexes: boundary
layers in CFD \cite{puso2006stabilized}, inaccuracy or locking problems in
solid mechanics \cite{benzley1995comparison}.

From a high order spectral finite element perspective, hex meshes provide
considerable advantages. First, although this is not specific to spectral
finite elements, a hex mesh contains about seven times
fewer elements than a tet mesh with the same number of vertices.
Fewer elements mean less data storage and a faster assembly procedure.
Taking advantage of the inherent tensor-product structure of hexahedral basis
functions one can dramatically reduce the number of floating point
operations for computing finite element operators. The local cartesian
structure of the mesh provides natural overlapping patches of 
elements that enables the construction of efficient local preconditioners. Finally,
spectral hex-meshes can achieve relatively high throughput on GPUs
following the approaches described below.

The use of GPUs for accelerating finite element solvers for elliptic problems
is of course not new. In early work G{\"o}ddeke  et al \cite{goddeke2007exploring} investigated scalability of
finite element solvers on GPU clusters. Later G{\"o}duke described  multigrid methods 
for finite element methods on GPU clusters \cite{goddeke2011fast}. Cecka et al \cite{cecka2011assembly}  and Markall et al \cite{markall2013finite}
discussed 
algorithms for efficient stiffness matrix assembly on GPUs. Knepley et al \cite{knepley2013finite} described algorithms
for efficient evaluation of finite element integrals on GPUs. Gaveled et al \cite{geveler2013towards} introduced
a finite element toolkit that integrates geometric multigrid techniques with sparse approximate inverse algorithms on GPUs. Furthermore, 
pushing the envelope of GPU based finite element software design Fu et al \cite{fu2014architecting} describe a systematic 
approach to pipelining finite element methods. Largely these prior approaches have focused on optimizing the process of
stiffness matrix assembly. The current work differs by first using a high-order finite element approach and secondly adopting a matrix-free approach that in its leanest form 
only requires storage for mesh vertex coordinates,  residual vector, solution  vector, load vector, and indexing arrays. 

In this paper, we propose a numerical scheme that allows us to solve
Poisson-like problems on unstructured all-hex meshes using the massive
multi-threading capacities of modern computer hardware. An extensible
multi-threading programming API is used as a common kernel language
\cite{medina2014occa} to try our numerical scheme on different
devices (GPU and CPU) and using different thread programming interfaces
(CUDA, OpenCL, and OpenMP).

This paper is structured as follows. In \S\ref{sec:1}, standard
properties of spectral finite elements are presented in brief.
The numerical method is presented in \S\ref{sec:pcg} and \S\ref{sec:2scale}: preconditioned
conjugate gradients are used for solving linear systems. A two-scale
additive Schwartz preconditioner is used for accelerating the
convergence.  Details of implementation are presented in
\S\ref{sec:impl} and results are presented in \S\ref{sec:res}.

\section{Spectral Finite Elements on Hexahedral Meshes \label{sec:1}}

Consider a domain $\Omega \in R^3$ with boundary $\Gamma=\Gamma_D \cup
\Gamma_N$
and the following model problem: find $u(x,y,z)$ that satisfies
\begin{eqnarray}
\cc u  - \nabla \cdot (
\kappa \nabla u ) = s~~&\mbox{on}&~~\Omega, \nonumber \\
u = u_0 ~~&\mbox{on}&~~\Gamma_D \\
{\partial u \over \partial n} = g ~~&\mbox{on}&~~\Gamma_N
\label{eq:pde}
\end{eqnarray}
where $\cc(x,y,z)>0$, $\kappa(x,y,z)>0$ and $s(x,y,z)$ is a given source term.
We further suppose that $s$, $u_0$ and $g$ satisfy the standard regularity
assumptions and, without loss of generality, that $u_0 = 0$.
A weak formulation of \eqref{eq:pde} is: find $u \in H_0^1(\Omega)$ that satisfies
\begin{equation}
\int_{\Omega}  \left[\kappa \nabla
u \cdot \nabla w + \cc u~w\right] ~dxdydz = \int_{\Omega} r~w ~dx dy dz ~~\forall w \in H_0^1(\Omega)
\label{eq:pdeweak}
\end{equation}
where $H^1_0(\Omega)=\{u \in H^1(\Omega),~u|_\Gamma = 0\}.$

\subsection{Interpolation}
Consider now a mesh constructed of unstructured hexahedra. On each hexahedron $e$, the finite element interpolation basis is a tensor products of one dimensional basis of $P_n$ that are the set of Lagrangian interpolants $\phi_j(t),~~j=0,\dots,n$ on the Gauss-Lobatto Legendre (GLL) quadrature points in the reference domain: $t_i \in [-1,+1], i = 0,...,n$,$\phi_j(t_i)=\delta_{ij}$ \cite{deville2002high}.

In the reference hexahedron $\xi,\eta,\zeta \in [-1,+1]$ of element $e$,  fields are interpolated as
\begin{equation}u^e(\xi,\eta,\zeta) = \sum\limits_{i=0}^n \sum\limits_{j=0}^n \sum\limits_{k=0}^n u_{ijk;e}
\phi_i(\xi) \phi_j(\eta) \phi_k(\zeta) \label{eq:interp}\end{equation}
where $u_{ijk;e}$ are the values of $u$ at the $(n+1)^3$ nodes of element $e$.  We define the derivation matrix $D$ following \cite{deville2002high} as
\begin{equation}
D_{ij} = \left.{d \phi_i \over dt}\right|_{t = t_j}. \label{eq:D}
\end{equation}

\subsection{Local and Global vectors}
Consider a mesh made of $N_E$ unstructured hexahedra with a total of $N$ GLL
nodes and a scalar field $u$ interpolated on the mesh. In the following,
two representations of $u$ will be used, one that is defined locally to one
element and a second that is defined globally on the mesh.
The local version of $u$ is denoted by
$$u_{ijk;e},~~0 \leq i,j,k \leq n,~~ 1 \leq e \leq N_E.$$
A global indexing of the GLL nodes is defined that associates a unique
number to every GLL node $\nod$ of
the mesh. The global version of $u$ is noted
$$u_\nod,~~1\leq \nod \leq N. $$
A local-to-global indexing table $\nod = \ix(ijk;e)$ provides
a global index $\nod$ for local node $(ijk;e)$.
A {\bf scatter} operation \cite{fischer2000overlapping} consists of building the local
representation of a vector from its global representation:
\begin{equation}u_{ijk;e} = u_{ \ix(ijk;e)}.\label{eq:scatter}\end{equation}
The {\bf gather} operator allows to
compute global vectors from local vectors.
It requires the definition of a global-to-local indexing table
$I_{g\rightarrow l}(\nod)$ that returns
the $N_\nod$ local nodes $(i_{\nod_l}j_{\nod_l}k_{\nod_l};e_{\nod_l})$
that are associated to a given global node $\nod$:
\begin{equation}\ix(I_{g\rightarrow l}(\nod)) =  \ix(i_{\nod_l}j_{\nod_l}k_{\nod_l};e_{\nod_l}) = \nod,~~l=1,\dots,N_\nod. \label{eq:gather}\end{equation}

\subsection{Geometry}
The unstructured nature of the hexahedral meshes that are considered
here force us to consider the geometry of each individual element $e$.
The geometry of element $e$ is defined through
its mapping
$$x^e(\xi,\eta,\zeta),~~y^e(\xi,\eta,\zeta),~~z^e(\xi,\eta,\zeta)$$
between the reference cube $\xi,\eta,\zeta \in [-1,+1]$ and
the element $e$. We define the Jacobian
$$J^e(\xi,\eta,\zeta) = \left[
\begin{array}{ccc} {\partial x^e \over \partial \xi} & {\partial x^e \over  \partial
    \eta} & {\partial x^e \over \partial \zeta} \\
{\partial y^e \over \partial \xi} & {\partial y^e \over \partial
    \eta} & {\partial y^e \over \partial \zeta}\\
{\partial z^e \over \partial \xi} & {\partial z^e \over \partial
    \eta} & {\partial z^e \over \partial \zeta}\end{array}\right],
$$
its determinant $|J|^e = \det{J^e}$ and the symmetric metric tensor
$G^e=(J^e)^T J^e$.

\subsection{Integration}
In the spectral element method formulation GLL points are both used for interpolation and integration purposes:
$$
\int_e f(x,y,z) dx dy dz \simeq \sum\limits_{i=0}^n
\sum\limits_{j=0}^n \sum\limits_{k=0}^n f_{ijk;e}\underbrace{\rho_i \rho_j \rho_k |J|_{ijk;e}}_{m_{ijk;e}}
$$
where the $\rho_j$'s are $1D$ integration weights.
GLL points are sub-optimal integration points: they only allow to
exactly integrate  a polynomial of order $2n-1$

The computation of \eqref{eq:pdeweak} is performed in two steps.
Local values of the residuals are computed at every GLL point of every element:
$$r_{ijk;e} = m_{ijk;e} \left[\kappa_{ijk;e} \left.\nabla
    u\right|_{ijk;e}  \cdot \nabla \left(\phi_i\phi_j
  \phi_k\right)|_{ijk;e} +
    c_{ijk;e} u_{ijk;e} \right].$$
Then local residuals $r_{ijk;e}$ are
{\bf gathered}  to global GLL nodes $\nod$ using \eqref{eq:gather} as
$$r_\nod = \sum\limits_{l=1}^{N_\nod} r_{i_\nod j_\nod k_\nod;e}.$$
Equations \eqref{eq:interp} and \eqref{eq:D} allow to compute
$$\left.{\partial u \over \partial \xi}\right|_{ijk;e} = \sum\limits_{m=0}^n D_{im} u_{mjk;e},~~
\left.{\partial u \over \partial \eta}\right|_{ijk;e}   = \sum\limits_{m=0}^n D_{jm} u_{imk;e},~~
\left.{\partial u \over \partial \zeta}\right|_{ijk;e} = \sum\limits_{m=0}^n D_{km} u_{ijm;e}.$$
Thus, $r_{ijk;e}$ is computed as
\begin{eqnarray}r_{ijk;e} &=& m_{ijk;e} \left[\right. \nonumber \\
                                      \sum\limits_{m=0}^n && \kappa_{mjk;e} ~D_{mi}
  \left[G^1_{mjk;e} \left.{\partial u \over \partial \xi}\right|_{mjk;e} +
          G^2_{mjk;e} \left.{\partial u \over \partial \eta}\right|_{mjk;e}  +
          G^3_{mjk;e} \left.{\partial u \over \partial \zeta}\right|_{mjk;e}
        \right] + \nonumber \\
  \sum\limits_{m=0}^n   &&\kappa_{imk;e}  ~D_{mj}
  \left[G^2_{imk;e} \left.{\partial u \over \partial \xi}\right|_{imk;e} +
          G^4_{imk;e} \left.{\partial u \over \partial \eta}\right|_{imk;e}  +
          G^5_{imk;e} \left.{\partial u \over \partial \zeta}\right|_{imk;e}
        \right] + \nonumber \\
  \sum\limits_{m=0}^n   &&\kappa_{ijm;e} ~D_{mk}
  \left[G^3_{ijm;e} \left.{\partial u \over \partial \xi}\right|_{ijm;e} +
          G^5_{ijm;e} \left.{\partial u \over \partial \eta}\right|_{ijm;e}  +
          G^6_{ijm;e} \left.{\partial u \over \partial \zeta}\right|_{ijm;e}
        \right] +\nonumber \\ &&
        c_{ijk;e} u_{ijk;e} \left.\right]. \label{eq:R}
\end{eqnarray}
\section{Preconditioned Conjugate Gradients  \label{sec:pcg}}
The aim now is to solve our problem, i.e. find $u$ solution of
$$r_\nod (u) = 0,~~1 \leq \nod \leq N.$$
For that, we use preconditioned conjugate gradients (PCG) because of the symmetric positive
definite nature of our problem.
Algorithm \ref{algo:standard} describes the PCG procedure.
In this description, $\pr$ a preconditioner.
\begin{algorithm}
\caption{Preconditioned Conjugate Gradients}\label{algo:standard}
$u_{0}=0$\\
$r_{0}= r(u_0)$\\
$z_{0}={\pr}(r_{0})$ \\
$p_{0}=z_{0}$\\
\For{$k=0,1,2,\ldots$}
   {\label{fork}
     $f  = r(p_k)$ \\
     $\alpha_k={{z_{k}^{\top}r_{k}}\over{{ p}_k^{\top}f}}$     \\
     $u_{k+1}=u_{k}+\alpha_kp_k$\\
     $r_{k+1}=r_{k}-\alpha_kf$\\
     $z_{k+1}=\pr r_{k+1}$\\
     $\beta_{k+1}={{z_{k+1}^{\top}r_{k+1} } \over { z_{k}^{\top}r_{k} } }$\\
    $ p_{k+1}=z_{k+1}+\beta_{k+1}p_{k}$\\
   }
\end{algorithm}
The two expensive steps that are computed at each
iteration are actually computing $r(p_k)$ and computing
$\pr r_{k+1}$.

\section{A two scale preconditioner \label{sec:2scale}}
We use here an additive Schwarz preconditioner $P$
using overlapping subdomains \cite{fischer1997overlapping} plus a coarse grid
projection operator.
Preconditioner $\pr$ is based on solving (i) a coarse problem
with low order ($n=1$) elements on the mesh  and
(ii) local problems on overlapping subdomains.
The contributions of the coarse and fine preconditioners are
subsequently added:
$$\pr = \pr^c+ \pr^f.$$
\subsection{Coarse grid preconditioner}
In short, the coarse grid preconditioner works as follows
$$ \pr^c r = V^T \pr_0 V r.$$
$V$ is a restriction operator that projects the
high order residual onto the coarse space.
In our implementation, $\pr_0$ consists of algebraic multigrid
cycles on the finite element matrix computed at
polynomial order $n=1$. The prolongation operator
$V^T$ is chosen as the transpose of $V$ in order to preserve the
symmetry of the problem.

The essential ingredient of this multilevel is the
\textit{restriction} operator.  In \cite{hillewaert2006hierarchic}
authors show that $L^2$ projection is a good choice for $V$. In that
specific case:
$$V = C m^{-1}$$
where $m$ is the fine scale mass matrix and where $C$ the
``mixed'' mass or correlation matrix. Note that $m$ is the lumped diagonal mass matrix
when using spectral finite elements: the $m_\nod$'s
are the global components of the local weights
$m_{ijk;e}$:
$$m_\nod = \sum\limits_{l=1}^{N_\nod} m_{i_\nod j_\nod k_\nod;e}.$$

The application of $\pr^c$ to the residual $r$ proceeds as:
\begin{equation} \pr^c  r = m^{-1} \overbrace{C^{T}\overbrace{\pr_{0} \underbrace{C \underbrace{m^{-1} r}_y}_R}^Z}^z.\label{eq:coarse}\end{equation}
Assume that capital letters indices $I$, $J$ and $K$ are
coarse indices: $I,J,K =0,1.$
Local correlation matrices are computed as
$$C_{ijk,IJK;e} = \int_e \phi_i \phi_j \phi_k \Phi_I \Phi_J \Phi_K  dx dy dz.$$
where the $\Phi$'s are coarse $1D$ shape functions.
Using fine scale GLL points for integration, we have
$$C_{IJK,ijk;e} \simeq
\Phi_I(t_i) \Phi_J (t_j) \Phi_K(t_k) ~m_{ijk;e}.$$
Vandermonde matrix
$$B_{IJK,ijk} = \Phi_I(t_i) \Phi_J (t_j) \Phi_K(t_k)$$
of size $8 \times n^3$ is computed once and stored.

Restriction of the global residual $r_\nod$ starts by scaling it by the inverted
mass matrix to form $y_\nod = r_\nod / m_\nod$ and to use \eqref{eq:scatter} to
form the local vector $y_{ijk;e}$.
Then, $y_{ijk;e}$ is used to form the local coarse residual (see \eqref{eq:coarse}):
$$R_{IJK;e} = \sum\limits_{i=0}^n \sum\limits_{j=0}^n \sum\limits_{k=0}^n
   B_{IJK,ijk}  ~  y_{ijk;e} ~m_{ijk;e}.
$$
Local coarse residual $R_{IJK;e}$ is gathered to the coarse grid nodes
to form the global coarse residual $R$. Coarse preconditioner
 $\pr_0$ is then applied to get the global vector $Z= P_0  R$.
Global vector $Z$ is then scattered to form $Z_{IJK;e}$.
We finally prolongate $Z_{IJK;e}$ as
$$z_{ijk;e} = \sum\limits_{I=0}^1 \sum\limits_{J=0}^1 \sum\limits_{K=0}^1
   B_{IJK,ijk}  ~  Z_{IJK;e} ~m_{ijk;e}.
$$
and scatter $z_{ijk;e}$ to form its global version $z_\nod$. Global vector
$y_\nod$ is finally scaled by $1/m_\nod$ to form the \emph{coarse scale
  correction} $\left.\pr^c r\right|_\nod = z_\nod/m_\nod$.
\subsubsection{Algebraic multigrid}
An aggregation based algebraic multigrid method is used to obtain an approximate solution of coarse grid system. At first, the coarse grid matrix is computed using finite elements with polynomial order $n=1$. Then, a hierarchy of matrices are constructed using  an unsmooth aggregation method \cite{gandham2014agmg}. An approximate solution for the coarse grid problem is obtained by the application of a recursive K-cycle \cite{notay2010aggregation} along with damped Jacobi smoothing at each level in the hierarchy as described in \cite{gandham2014agmg}. Experimental results show that the algebraic multigrid method provides  h-independence convergence.     
\subsection{Local preconditioner}
\begin{figure}
\begin{center}
\includegraphics[width=.4\textwidth]{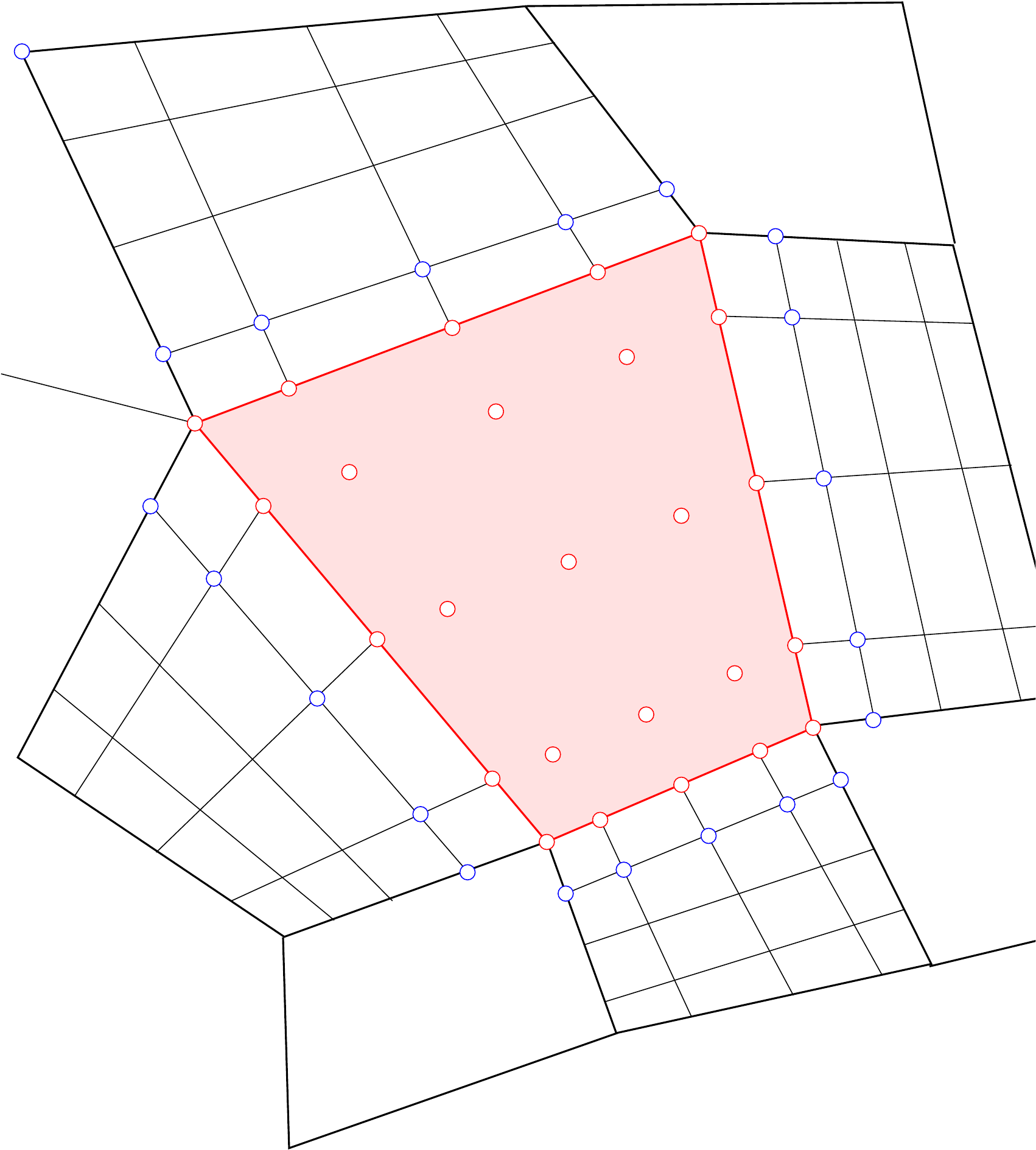}
\includegraphics[width=.5\textwidth]{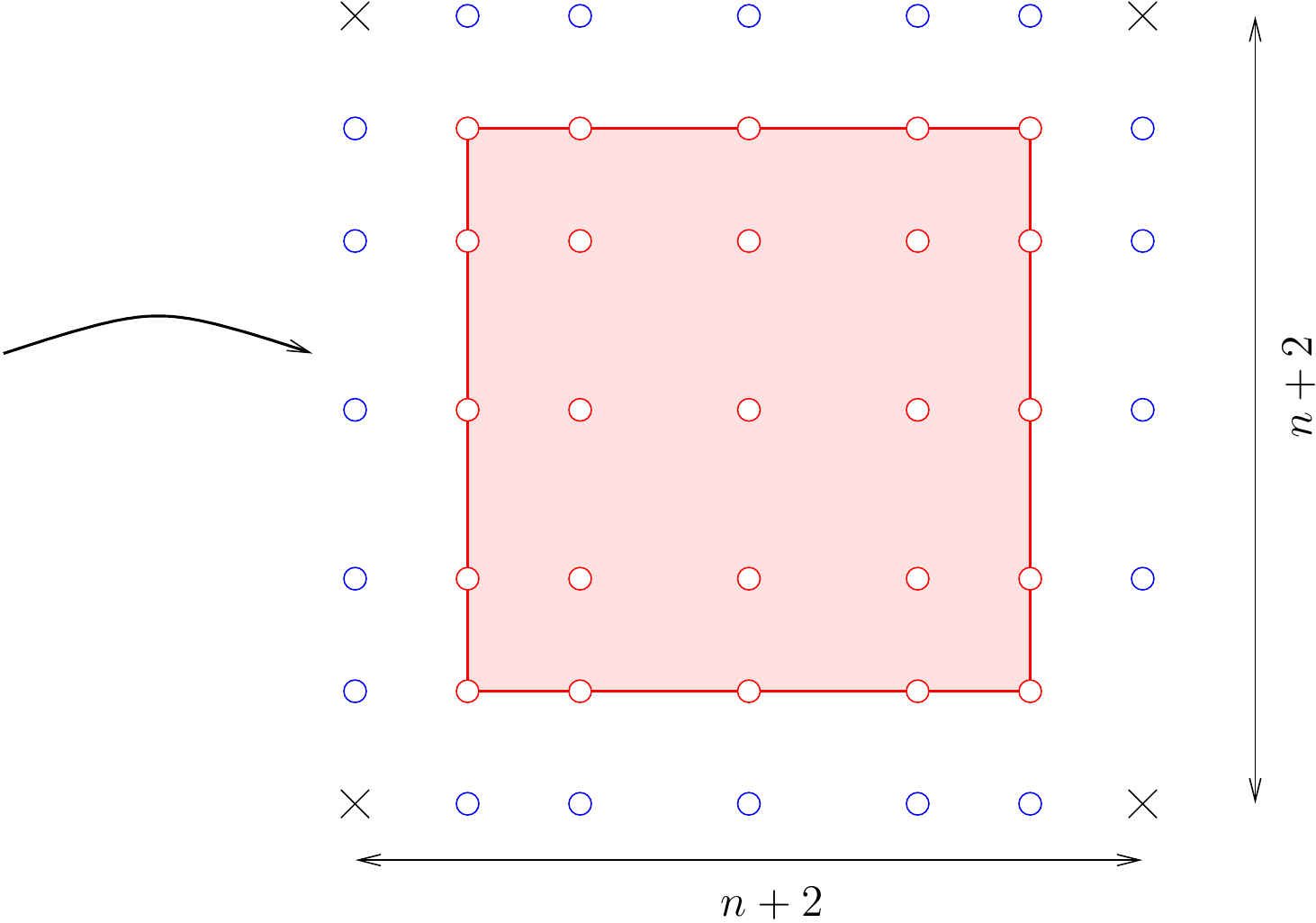}
\end{center}
\caption{Fine scale stencil for the local preconditioner. Left: GLL nodes shown superimposed on physical elements. Right: topological relationship of GLL nodes on reference element. \label{fig:ref}}
\end{figure}

An overlapping additive Schwartz approach is used to build the local
preconditioner. A set of $N_E$ overlapping subdomains is defined.
Each subdomain $s$ is an element of the mesh with one layer of nodes
overlapping neighbor elements on all its faces  (Figure \ref{fig:ref}), with a total of
$(n+3)^3$ GLL nodes per subdomain.

The restriction of $r_\nod$ on subdomain $s$ is denoted by $r_{ijk;s}$, $-1
\leq i,j,k \leq n+1$.  It is computed using a table $\ixx(ijk;s)$
 that allows to scatter global
vectors to  the vertices of the subdomains:
 $$r_{ijk;s} = r_{ \ixx(ijk;s)}.$$
The principle is to solve problem \eqref{eq:pdeweak} on each subdomain
i.e. find the \emph{fine scale correction} $z_{ijk;b}$ that is the solution
of \eqref{eq:pdeweak} with $r_{ijk;s} $ as right hand side and with
suitable boundary conditions.

We make an assumption that greatly
simplifies the computation of fine scale corrections, especially for
the case of unstructured meshes.
Fine scale corrections are computed without taking into account
geometric factors i.e. assuming that hexahedra's faces and edges
are aligned with the axis of coordinates $x,y$ and $z$. Each hexahedra
is approximated as a parallelogram of dimensions $h_x$, $h_y$ and
$h_z$.
This geometric simplification actually allows to compute coarse
scale corrections in an efficient way.

Consider a 1D pencil of size $h_x$ with
$$u(t) = \sum\limits_{i=0}^n u_i \phi_i(t) $$
with $x =  {t+1 \over 2} h_x$ and $u_i = u(t_i)$.
Assuming $\kappa$ and $\cc$ constant per element,
the 1D element matrix corresponding to  \eqref{eq:pdeweak} is
\begin{eqnarray}
K_{ij} = \int_0^{h_x} \left(\kappa {d \phi_i \over dx} {d \phi_j \over dx} +
\cc \phi_i \phi_j \right)dx
\simeq {2 \kappa \over h_x} \underbrace{\sum\limits_{m=0}^n \left (D_{im} D_{jm}  \rho_m\right)}_{d_{ij}}
  + {\cc h_x \over 2} \rho_j \delta_{ij}.
\label{eq:subd}
\end{eqnarray}
where $d_{ij}$ is the discrete second derivative.

\begin{figure}
\begin{center}
\includegraphics[width=.85\textwidth]{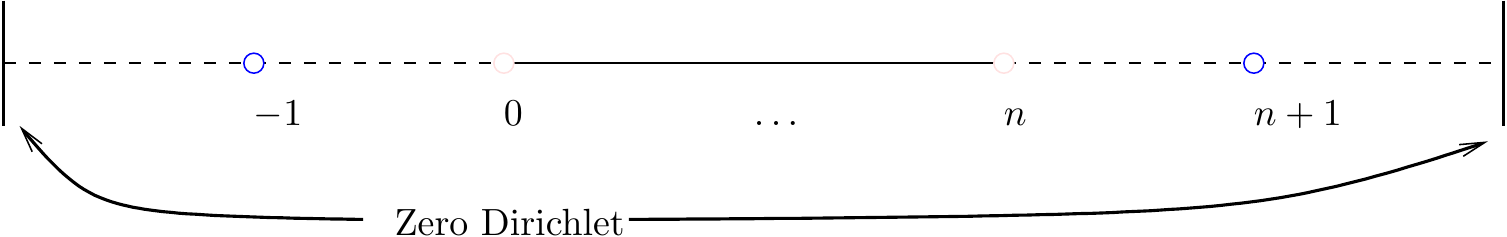}
\end{center}
\caption{The 1D pencil.\label{fig:1d}}
\end{figure}

Let's extended the pencil with one GLL vertex on
its left (point $-1$ on Figure \ref{fig:1d}) and on its right (point
$n+1$ on Figure \ref{fig:1d}).
The stiffness operator on the extended domain is computed using standard finite element
assembly procedure i.e. adding contribution of vertex $-1$ and $n+1$
in the finite element matrix. Homogeneous Dirichlet
boundary conditions are computed on points $-2$ and $n+2$ so that the
final matrix $l_{ij}$, $-1\leq i,j \leq n+1$ is invertible.
The 1D finite element problem is finally written  as
\begin{equation}
\tiny
{
\left(
{4\kappa \over  h^2_x}
\left[
\begin{array}{ccccc}
d_{11} &   d_{01} &        &  & \\
d_{10} & 2d_{00} & \hdots & d_{0n} & \\
 & \vdots & \ddots & \vdots &  \\
 & d_{n0} & \hdots & 2d_{00}  & d_{01} \\
&  &  & d_{10}  & d_{11} \\
\end{array}
\right]
+
{\cc}
\left[
\begin{array}{ccccc}
\rho_{1} &    &        &  & \\
 & 2\rho_{0} &  &  & \\
 &  & \ddots & &  \\
 &  &  & 2\rho_{0}  &  \\
&  &  &  & \rho_{1} \\
\end{array}
\right]\right)
\left[
\begin{array}{c}
z_{-1} \\ z_0 \\ \vdots \\ z_n \\ z_{n+1}
\end{array}
\right]
=
{2 \over h_x}
\left[
\begin{array}{c}
r_{-1} \\ r_0 \\ \vdots \\ r_n \\ r_{n+1}
\end{array}
\right]
}
\nonumber
\end{equation}
or in matrix form
$$
\left ({4\kappa \over  h^2_x} K + c M\right) z =  {2 \over  h_x}
r ~~\rightarrow~~~\left ({4\kappa \over  h^2_x} M^{-1} K + c I\right) z =  {2 \over  h_x}
M^{-1}r.$$
Matrix $L = M^{-1} K$ is diagonalizable
$$L = V^{-1} \Lambda V$$ and has real and positive eigenvalues $\lambda_i$.
The diagonalization of $L$ leads to the explicit solution:
$$
z =  \left[V^{-1}\left ({4\kappa \over  h^2_x} \Lambda  + c I \right)^{-1} V\right] {2 \over  h_x}
 M^{-1}r.
$$
In 3D,  the local subproblem consist in finding the fine scale correction
$z_{ijk;b}$ solution of
$$4\kappa \sum\limits_{m=-1}^{n+1}
       \left({1 \over  h_x^2}L_{im}  z_{mjk;s} +
               {1 \over h_y^2} L_{jm}  z_{imk;s} +
               {1 \over h_z^2} L_{km} z_{ijm;s} \right) + c
             z_{ijk;s} = r'_{ijk;s}$$
with
$$r'_{ijk;s}= { 8 \over h_x h_y h_z} {1 \over M_{ii} M_{jj}M_{kk}}r_{ijk;s}.$$
It is then possible to compute the solution as we did in 1D:
\begin{eqnarray}
\label{eq:z}
z_{ijk;s}
=
\sum\limits_{d=-1}^{n+1}
\sum\limits_{e=-1}^{n+1}
\sum\limits_{f=-1}^{n+1}
V^{-1}_{id} V^{-1}_{je} V^{-1}_{fk}
~{ 1 \over 4 \kappa
    \left({\lambda_d \over h_x^2} + {\lambda_e \over
          h_y^2} + {\lambda_f \over  h_z^2} \right) + c} \nonumber \\
\sum\limits_{a=-1}^{n+1} \sum\limits_{b=-1}^{n+1} \sum\limits_{c=-1}^{n+1}  V_{ad} V_{be}
V_{cf} r'_{abc;s}.
\end{eqnarray}
The fine correction for subdomain $s$ is added to the fine preconditioner
$$\left. \pr^f \right|_{\ixx(ijk;s)}  \leftarrow  \left. \pr^f
\right|_{\ixx(ijk;s)} + z_{ijk;s}, ~~i,j,k \in [-1,n+1]^3.
$$
\section{Implementation \label{sec:impl}}
Our implementation uses OCCA which is a novel approach that includes a
unified kernel language that expands to multiple different threading languages \cite{medina2014occa}
that has recently been demonstrated to be an effective platform for implementing
discretizations of hyperbolic problems
\cite{medina2014high, gandham2014gpu, modave2015accelerated}. The two dominant cost
kernels are the computation of residual $r_{ijk;e}$ and the
fine grid preconditioner $z_{ijk;s}$ (see \eqref{eq:z}).

General purpose parallel programming on GPUs allow a two-level
parallelism. GPU devices have a large number of multiprocessing units
that can run threads. Threads are organized in blocks and whole thread blocks
are executed by a multiprocessing unit. The global device memory of the GPU
is accessible by all the threads of all blocks. Some memory is only
accessible by a given block of threads (shared memory) and has relatively low latency compared to global device memory.
In our implementation, each element is assigned to a thread block and
residual calculations at each GLL node are assigned to one thread.

\subsection{An OCCA kernel for computing $r_{ijk;e}$}
Algorithm \ref{algo:R} describes our implementation for computing
$r_{ijk;e}$: variables with superscript $s$ implies that the data is
stored in on-chip shared memory accessible to the threads processing
element $e$.

Parameters $\kappa$ and $c$ that are element-based are copied in
shared memory, as well as
the derivative matrix $D_{ij}$ and the current solution  $u$.
\begin{algorithm}
\caption{Computation of $r_{ijk;e}$}\label{algo:R}
{
\For{$e=1,\ldots,N_E$ }
   {
      $\kappa_e^s = \kappa_e,~~c_e^s = c_e.$\\
      \For{$i,j \in [0,n]^2$}{
          $D_{ij}^s = D_{ij}.$\\
      }
      \For{$i,j,k \in [0,n]^3$}{
          $u_{ijk;e}^s = u_{ \ix(ijk;e)}.$\\
      }
      {{\tt barrier();}}\\
      \For{$i,j,k \in [0,n]^3$}{
      $\left.{\partial u \over \partial \xi}\right|_{ijk;e} = \sum\limits_{m=0}^n D^s_{im} u_{mjk;e}^s$,
       $\left.{\partial u \over \partial \eta}\right|_{ijk;e}   = \sum\limits_{m=0}^n D^s_{jm} u_{imk;e}^s$,
       $\left.{\partial u \over \partial \zeta}\right|_{ijk;e} =
       \sum\limits_{m=0}^n D^s_{km} u_{ijm;e}^s.$\\
      {Compute $A_{ijk;e}^s$, $B_{ijk;e}^s$, $C_{ijk;e}^s.$ as in \eqref{eq:ABC}}\\
      }
      {{\tt barrier();}}\\
      \For{$i,j,k \in [0,n]^3$}{
     {
       \begin{eqnarray}r_{ijk;e} = \kappa^s_{e} \sum\limits_{m=0}^n \left[D_{mi}^s
           A_{mjk;e}^s + D_{mj}^s B_{imk;e}^s+ D_{mk}^s
           C_{ijm;e}^s\right]+ c^s_{e} u_{ijk;e}^s  m_{ijk;e}.  \nonumber
       \end{eqnarray}
      }   \\
   }
}
}
\end{algorithm}
Then, $(n+1)^3$
threads are launched in order to compute
(i) derivatives of $u$ ($2 \times 3 \times (n+1)$ operations per thread),
(ii) gradients
\begin{eqnarray}A_{ijk:e}^s &=&
  \left[G^1_{ijk;e} \left.{\partial u \over \partial \xi}\right|_{ijk;e} +
          G^2_{ijk;e} \left.{\partial u \over \partial \eta}\right|_{ijk;e}  +
          G^3_{ijk;e} \left.{\partial u \over \partial \zeta}\right|_{ijk;e}
        \right], \nonumber \\
B_{ijk:e}^s &=&
  \left[G^2_{ijk;e} \left.{\partial u \over \partial \xi}\right|_{ijk;e} +
          G^4_{ijk;e} \left.{\partial u \over \partial \eta}\right|_{ijk;e}  +
          G^5_{ijk;e} \left.{\partial u \over \partial \zeta}\right|_{ijk;e}
        \right],\nonumber \\
C_{ijk:e}^s &=&
  \left[G^3_{ijk;e} \left.{\partial u \over \partial \xi}\right|_{ijk;e} +
          G^5_{ijk;e} \left.{\partial u \over \partial \eta}\right|_{ijk;e}  +
          G^6_{ijk;e} \left.{\partial u \over \partial \zeta}\right|_{ijk;e}
        \right].\label{eq:ABC}
\end{eqnarray}
($3 \times 5$ operations per thread). Finally,  the residual is computed
at every GLL node ($2 \times 3 \times (n+1) + 3$ operations per thread). Two barriers
separate the three main parts of the algorithm.
The maximum number of threads per thread block in CUDA is currently 1024,
our kernel can only be used up to $n=9$ for which $(n+1)^3=1000.$ In OpenCL the number of work-items per work-group is limited to 256 on AMD GPUs and consequently we are limited to $n=5$ in that case.
The total number of operations $O_R$ for computing $r_{ijk;e}$ is
$$O_R = N_E \times \left[ 12 \times (n+1)^4 + 18 \times (n+1)^3\right].$$
Memory bandwidth quantifies the rate of throughput of data transfer between GPU and the GPU global device memory.
In modern GPU architectures, a
typical value for the memory bandwidth is 250 GB/sec i.e. about
75 billion floats per second.
The total number of floating point numbers transferred from the global
memory to the shared memory here is
$$B_R =N_E \times \left[ 10 \times (n+1)^3 + (n+1)^2 + 2\right]$$
where the dominant term  $10 \times N_E \times (n+1)^3 $
correspond (per GLL point) to the $7$ geometric factors
$G^m_{ijk;e}, 1 \leq m\leq 6$ and the weighted masses $m_{ijk;e}$,
to the solution $u_{ijk;e}$,
to the residual itself $r_{ijk;e}$ and to the table $\ix(ijk;e)$ that is used
for scattering $u_\nod$.

At this point, it is interesting to look at the ratio
$O_R/B_R$ Any modern GPU can theoretically
deliver over one TeraFlop while its bandwidth is limited to 250
GB/sec. This means that the value  $O_R/B_R$ should be over
$20$ to ensure that the computation is not limited by bandwidth
limitations.  Here, ratio  $O_R/B_R$  is equal to
seven for $n=4$: we expect the performances of our kernel to be
strongly limited by the GPU bandwidth, especially at low orders.
It is therefore interesting to consider an alternative approach where the
seven geometric factors are computed on the fly. In this case, the only
geometric data that have to be shipped to the GPU are the position of the
vertices of the mesh  as well as the connectivity table of the
hexahedra. More operations are
required to compute the geometric factors: more precisely,
$242$ floating point operations per GLL node are required to compute
all seven geometric factors. In what follows, $O_R$ will be computed in
two different manners, whether the extra
computations that have been done for computing geometric factors are
or are not not taken into account. In this new
scheme, memory bandwidth is indeed reduced to
$$B_R =N_E \times \left[ 3 \times (n+1)^3 + (n+1)^2 + 2\right]$$

\subsection{An OCCA kernel for computing $z_{ijk;s}$}
The local preconditioning step that computes$z_{ijk;s}$ is the remaining expensive part of the
algorithm. Each subdomain $s$ is assigned to a thread block. Then,
$(n+3)^3$ threads are launched in every block in order to compute
$z_{ijk;s}$, which limits the order to $n=7$ for CUDA and $n=3$ for OpenCL.
Equation \eqref{eq:z} explicitly provides a formula for $p_{ijk;s}$.
The most significant computations are the
six successive products with the left and right eigenvectors
$V^{-1}$ and $V$  of the 1D operator.
More precisely, the total number of operations $O_P$ for computing $p_{ijk;s}$ is
$$O_P =  N_E \times \left[ 6 \times (n+3)^4 + 15 \times (n+3)^3\right].$$

Geometric factors are not loaded by this kernel, which implies that the
total amount of data that is transferred is reduced. Vectors
$z_{ijk;b}$, $r_{ijk;b}$ as well as the scattering table
$\ixx(ijk;b)$ are the three large local vectors that have to be transferred
from the global memory to the shared memory.
$$B_P= 4 \times N_E \times \left[ 3 \times (n+3)^3 + 4 \times (n+3)^2\right].$$
The ratio $O_P/B_P$ is already large at low orders to expect good
performances of the kernel for any $n$.

\subsection{A strategy for computing the preconditioning step}
We have shown that the most intensive part of the computation is the
evaluation of the local preconditioner  $z_{ijk;s}$. In an additive
Schwartz procedure, coarse and fine preconditioners can be computed
independently and summed afterwards.
Here, we propose to compute the coarse scale
preconditioner on the CPU while computing the fine scale part on the
GPU. At low orders, we expect that the computation of the coarse
scale preconditioner on the CPU will hide the computation of the fine
scale preconditioner on the GPU. At higher order, the fine scale
preconditioner will be more expensive and will hide the computation of
the coarse scale part. There may exist a sweet spot where both are
taking equivalent time.
\section{Results \label{sec:res}}
\subsection{Poisson problem}
We choose the parameters in out model problem \eqref{eq:pde} in order to
turn it into a Poisson equation.  We choose $\kappa = 1$,
$\cc = 0$ and $s=1$.
\subsubsection{Analysis of the two scale strategy}
\begin{figure}
\begin{center}
\begin{tabular} {ccc}
\includegraphics[width=.3\textwidth]{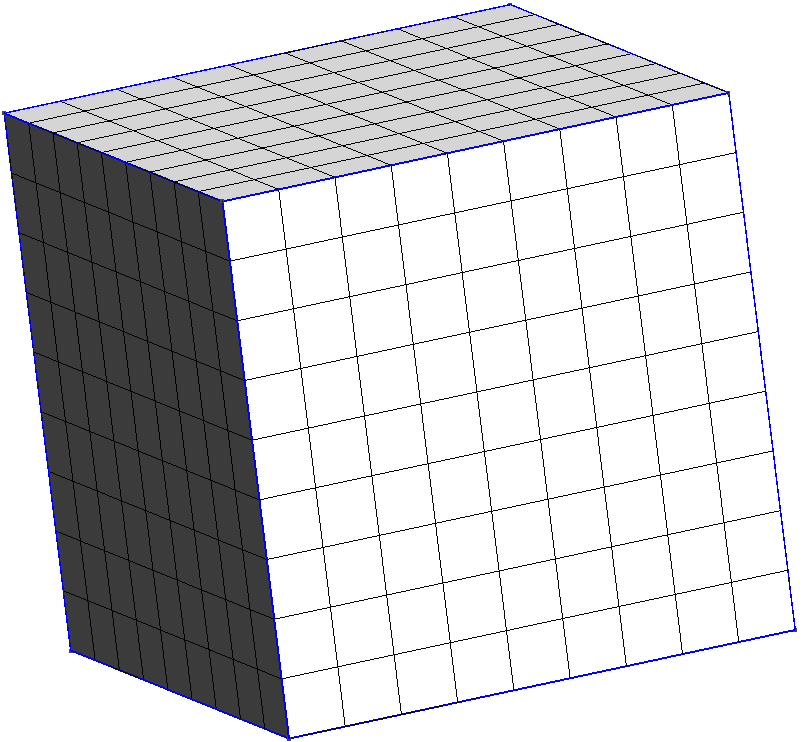}&
\includegraphics[width=.3\textwidth]{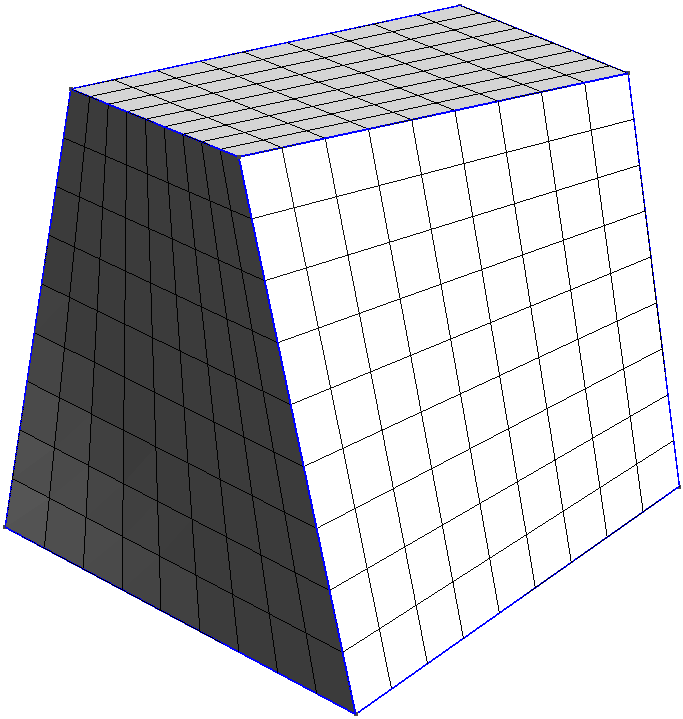}&
\includegraphics[width=.3\textwidth]{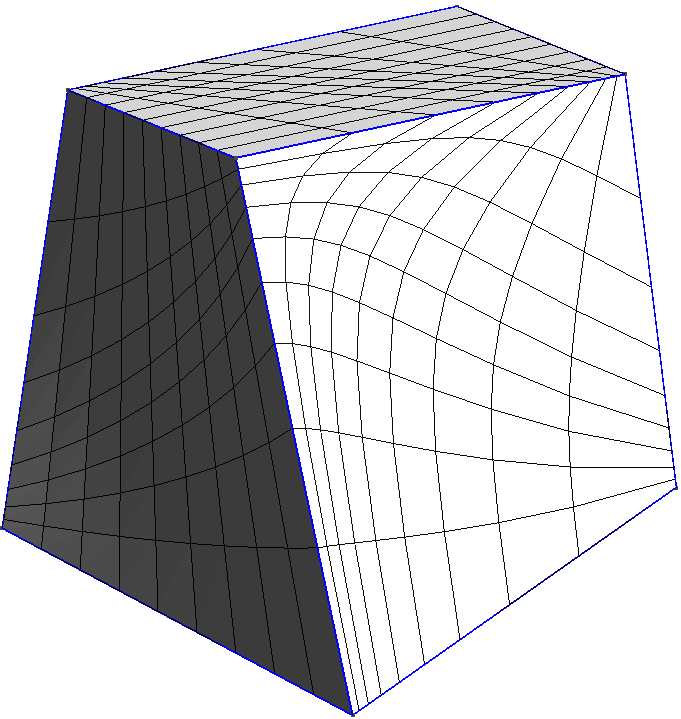}\\
Mesh 1 & Mesh 2 & Mesh3
\end{tabular}
\end{center}
\caption{Three meshes. Left: cube domain meshed with uniform elements. Center: nearly uniform elements on distorted cube domain. Right: strongly distorted elements forming mesh for distorted cube.\label{fig:meshes}}
\end{figure}
 In this section, we investigate the proposed numerical scheme. Three coarse meshes with $N_E=8^3$ hexes are considered (see Figure
\ref{fig:meshes}) and uniformly refined up to $N_E=128^3$. A
polynomial order $n=3$ is used for the computations. The strategy that
has been chosen for preconditioning is $h$-optimal: the number
of CG iterations is asymptotically stable for all meshes. If only the fine
scale preconditioner is used, the number of iterations doubles
when the mesh size is divided by two. It must be noted that the computation
of $\pr^c $ (on the CPU) is fully hidden by the computation of $\pr^f$
(on the GPU) for $n>2$. The coarse scale preconditioner has a zero
cost for a large benefit.

Note that the number of CG
iterations increases with polynomial order: for mesh $2$ at level of
refinement $32^3$ and for $n=7$, the number of CG iterations grows to
$23$. This is due to the fact that the size of the overlap decreases
while $n$ grows. A larger overlap would fix that issue \cite{pavarino1993additive}.
\begin{table}
\begin{center}
\begin{tabular}{c|ccccc}
\hline
DOF count          &  $8^3$          & $16^3$       & $32^3$        & $64^3$ &  $128^3$ \\
$N$   &  $1.5~10^4$ & $1.1~10^5$& $9.1~10^5$ & $7.1~10^6$ & $5.7~10^7$\\
\hline
 & \multicolumn{5}{c}{Two scale strategy: $\pr = \pr^c + \pr^f$}\\
\hline
Mesh 1   & 10    & 11      & 13   & 13 &  13\\
Mesh 2   & 10    & 13      & 15   & 15 &  15\\
Mesh 3   & 12    & 18      & 21   & 21 &  21\\
\hline
 & \multicolumn{5}{c}{Fine scale preconditioner only $\pr = \pr^f$}\\
\hline
Mesh 1   & 10    & 16      & 28   & 52 &  103\\
Mesh 2   & 13    & 20      & 32   & 60 &  118\\
Mesh 3   & 14    & 23     & 41   & 94 &  185\\
\hline
\end{tabular}
\end{center}
\caption{Number of Conjugate Gradient iterations as a function of the
  preconditioning strategy.
\label{tab:poisson2}}
\end{table}

\subsubsection{Performances of the kernels}
\begin{table}
\begin{center}
{\small
\begin{tabular}{r|rrrrrrr}
\hline
 & $n=2$ & $n=3$ & $n=4$ & $n=5$ & $n=6$ & $n=7$ \\
\hline
$N$  & $2.7~10^5$ & $9.1~10^5$ & $2.1~10^6$&$4.1~10^6$ & $7.1~10^6$ & $1.1~10^7$\\
\hline
$r_{ijk;e}$ (GFLOPs)    & $70$&  $145$ &  $227$ & $289$ &  $325$ & $362$\\
$r'_{ijk;e}$ (GFLOPs)   & $75/336$ &  $135/763$  &  $250/1050$    & $272/1020$ & $286/1050$ & $293/1050$\\
$z_{ijk;s}$(GFLOPs)    & $370$ &  $371$ &  $408$&  $453$ & $421$ & $373$\\
\hline
$r_{ijk;e}$ (GB/sec)& $59$ & $101$& $135$ & $150$ & $150$ & $150$\\
$r'_{ijk;e}$ (GB/sec)& $32$ & $48$& $75$ & $71$ & $66$ & $61$\\
$z_{ijk;s}$(GB/sec) & $96$ & $80$ & $76$ &  $74$ &  $61$ &   $49$\\
\hline
Wall Time  (sec)  & $0.10$ &$0.12$ &$0.23$ & $0.34$ & $0.53$& $0.83$\\
Wall Time' (sec)  & $0.11$ &$0.12$ &$0.22$ & $0.34$ & $0.52$ & $0.80$\\
\hline
\end{tabular}
}
\end{center}
\caption{ Performances of the two expensive kernels
$r_{ijk;e}$ and $z_{ijk;s}$ on a NVIDIA GTX 980 GPU using
OpenCL. A uniform mesh of $32^3$ hexahedra was used for the
test. Apostrophes indicate that geometric factors were computed on the
fly. In row relative to $r'_{ijk;e}$, the two numbers correspond
whether the cost cof computing geometrical factors was not or was taken
into account in $O_R$.
\label{tab:poisson4}}
\end{table}
We consider here the $32^3$ version of Mesh 2 (Figure \ref{fig:meshes}) .
Table \ref{tab:poisson4} present flops counts and bandwidth results for
different polynomial orders for a uniform mesh of
$32^3$ elements. Quantities with a prime refer to the kernels for
which the geometric factors were computed on the fly.
In Table \ref{tab:poisson4}, two numbers are given in the row relative to
$r'_{ijk;e}$'s flop count. The first one does not take into account the operations
that are required to compute the geometry factors while the second
number adds to the flop count $N_E \times \left[ 242 \times
  (n+1)^3\right]$ floating point operations which correspond to the
on the fly computation of the $7$ geometric factors.

It is interesting to see that the difference in wall clock time
between the two approaches is below measurement error.
Computing
geometric factors in an element is relatively floating point intensive: it consist essentially of loading the
$24$ coordinates of the eight vertices on registers and performing $242$
floating point operations required to compute the geometric factors $G_{ijk;e^m}$, $1\leq m \leq
6$ and $m_{ijk;e}$ that are themselves stored on registers. The specific GPU that has been used has a
very high peak floating point performance (over five teraflops in single precision). Computing geometric factors
is done a a rate that is close to the peak performance of the machine
while computing the rest of the kernel $r'_{ijk;e}$ requires memory access
that are bounded by bandwidth. In the case of  kernel $r_{ijk;e}$, loading geometric
factors requires bandwidth: the GPU stalls quickly at about $150$
GB/sec and so does the flop count. If geometric factors are taken
into account in the flop count, the performances of this kernel climb
to over one teraflop!

We have done the same computations on  a different GPU
(see Table \ref{tab:poisson} ) with a lower throughput. Kernel $r'_{ijk;e}$
is slightly slower than  kernel $r_{ijk;e}$.
\begin{table}
\begin{center}
\begin{tabular}{c|cccccc}
\hline
 & $n=2$ & $n=3$ & $n=4$ & $n=5$ & $n=6$ & $n=7$ \\
\hline
$N$  & $2.7~10^5$ & $9.1~10^5$ & $2.1~10^6$&$4.1~10^6$ & $7.1~10^6$ & $1.1~10^7$\\
\hline
$r_{ijk;e}$ (GFLOPs)    & $45$&  $117$ &  $194$ & $234$ &  $227$ & $287$\\
$r'_{ijk;e}$ (GFLOPs)   & $28$ &  $68$  &  $91$    & $106$ & $105$ & $112$\\
$z_{ijk;s}$(GFLOPs)    & $313$ &  $296$ &  $285$&  $380$ & $276$ & $371$\\
\hline
$r_{ijk;e}$ (GB/sec)& $38$ & $82$& $115$ & $121$ & $104$ & $118$\\
$r'_{ijk;e}$ (GB/sec)& $12$ & $24$& $27$ & $28$ & $24$ & $23$\\
$z_{ijk;s}$(GB/sec) & $81$ & $64$ & $53$ &  $62$ &  $40$ &   $48$\\
\hline
Wall Time  (sec)  & $0.10$ &$0.13$ &$0.26$ & $0.39$ & $0.67$& $0.92$\\
Wall Time' (sec)  & $0.11$ &$0.14$ &$0.29$ & $0.46$ & $0.79$ & $1.13$\\
\hline
\end{tabular}
\end{center}
\caption{ Performances of the two dominant cost kernels
$r_{ijk;e}$ and $z_{ijk;s}$ on a NVIDIA K40 GPU using
CUDA. A uniform mesh of $32^3$ hexahedra was used for the test.
Apostrophes indicate that geometric factors were computed on the
fly.
\label{tab:poisson}}
\end{table}


\subsection{Heat equation}
\begin{figure}
\begin{center}
\includegraphics[width=.92\textwidth]{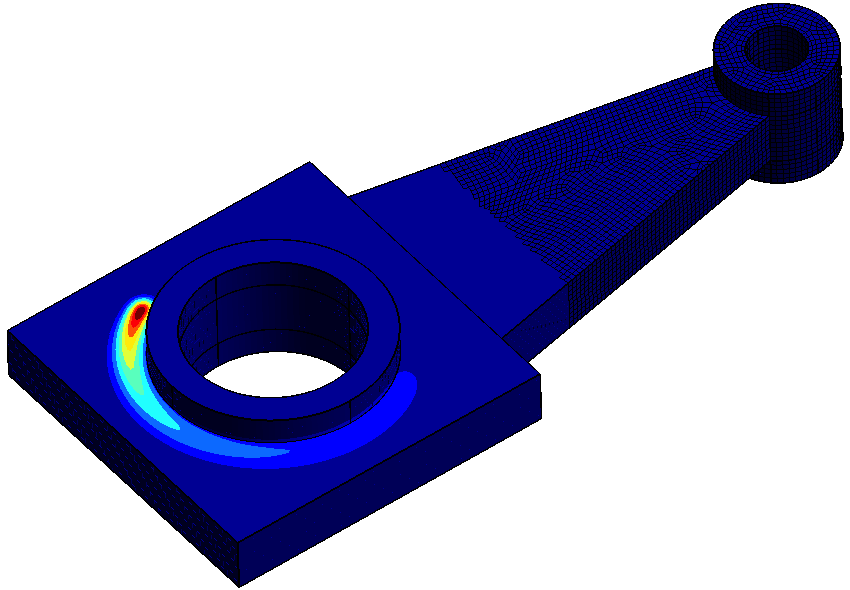}
\end{center}
\caption{Rod geometry considered for the heat equation with the
  hex-mesh of $62540$ hexahedra.\label{fig:rod}. Temperature field is
  shown at iteration $70$ which correspond to $t=2.8$ seconds. The
  motion of the heat source is visible on the plot. Mesh is visible on
  the right part of the plot\label{fig:rod}}
\end{figure}
We choose the parameters in out model problem \eqref{eq:pde} in order to
turn it into a heat equation.
A backward Euler scheme is used to
advance in time.  We choose $\kappa = 10^{-2} [m^2/\mbox{sec}]$ (aluminium),
a time step of $\Delta t= 0.04$ seconds and $\cc = 1/\Delta t$. The source term in
\eqref{eq:pde} is chosen as $r=u_\nod^{t-\Delta t}/\Delta t + Q / (\rho c_p)$ where
$u_\nod^{t-\Delta t}$ is the solution at previous time step, $Q = 1000~[W]$ is a
volume heat source $\rho = 7000~ [kg/m^3]$ and $c_P=0.8$. The volumetric
heat source $Q(x,y,z)$ is moved while time is advancing (see Figure \ref{fig:rod}).

The geometry and the mesh that we consider are presented in Figure \ref{fig:rod}.  
We solved $70$ time steps  of the discretized heat equation on this mesh using
different devices (GPU and CPU) and using
different threading systems (CUDA, OpenCL, and OpenMP).
All computations are performed in single precision arithmetic, either on an
NVIDIA $\copyright$ GTX 980 GPU or on a 8 core Intel $\copyright$ Core$\texttrademark$ i7-5960X CPU @
3.00GHz that has a theoretical peak performance of about $300$ GFLOPS.
Results are compiled in Table \ref{tab:perf}. Geometrical factors were
always computed on the fly and flop count takes into account their
computation.
\begin{table}
\begin{center}
{\small
\begin{tabular}{l|rrrrrrr}
\hline
 & $n=2$ & $n=3$ & $n=4$ & $n=5$ & $n=6$ & $n=7$ \\
\hline
device mem.(MB) & $178$ & $370$ &$587$&$1147$ & $1796$ & $2660$\\
$N$ & $5.3~10^5$ & $1.7~10^6$ &$4.1~10^6$&$8.0~10^6$ & $13.8~10^6$ & $21.9~10^6$\\
\hline
 & \multicolumn{5}{c}{Total wall time (sec) for 70 time steps}\\
\hline
CUDA(GPU)                & $13.4$ & $14.8$& $32.1$ & $53.6$ & $90.3$ & $134.9$\\
OpenCL(GPU)             & $12.2$ & $13.9$& $31.8$ & $49.9$ & $83.0$ & $126.3$\\
OpenMP(16 threads)  & $100.7$ & $127.4$& $355.1$ & $509.9$ & $901.5$ & $1393.5$\\
OpenCL(16 threads)  & $102.7$ & $156.2$& $384.2$ & {\color{red} $479.7$} & $1052.1$ & {\color{red}$ 1168.5$}\\
\hline
 & \multicolumn{5}{c}{Performance in GFLOPs of $r'_{ijk;e}$}\\
\hline
CUDA(GPU)                 & $674.0$ & $923.0$& $926.0$ & $874.0$ & $806.0$ &$795.0$\\
OpenCL(GPU)              & $1140.0$ & $1660.0$& $1690.0$ & $1390.0$ & $1380.0$ &$1310.0$\\
OpenMP(16 threads)   & $24.5$ & $68.8$& $59.3$ & $58.5$ & $57.8$ &$90.0$\\
OpenCL(16 threads)   & $30.2$ & $39.3$& $42.4$ & $42.9$ & $43.0$ &{\color{red} $206.0$}\\
\hline
 & \multicolumn{5}{c}{Performance in GFLOPs of $z_{ijk;e}$}\\
\hline
CUDA(GPU)                 & $479.0$ & $533.0$& $492.0$ & $535.0$ & $479.0$ &$591.0$\\
OpenCL(GPU)              & $422.0$ & $385.0$& $457.0$ & $470.0$ & $445.0$ &$478.0$\\
OpenMP(16 threads)   & $27.9$ & $40.4$& $38.9$ & $63.9$& $49.4$ & $51.5$\\
OpenCL(16 threads)   & $23.7$ & $26.6$& $27.4$ & {\color{red}$82.6$} & $28.8$ & $28.5$\\
\hline
 & \multicolumn{5}{c}{Bandwidth in GB/sec for $r'_{ijk;e}$}\\
\hline
CUDA(GPU)                 & $56.9$ & $74.0$& $70.9$ & $64.2$ & $56.9$ &$54.1$\\
OpenCL(GPU)              & $96.1$ & $133.0$& $129.0$ & $102.0$ & $97.7$ &$89.9$\\
OpenMP(16 threads)   & $2.1$ & $5.5$& $4.5$ & $4.3$ & $4.1$ & $6.1$\\
OpenCL(16 threads)   & $2.5$ & $3.1$& $3.2$ & $3.1$ & $3.0$ &
{\color{red} $14.0$}\\
\hline
 & \multicolumn{5}{c}{Bandwidth in GB/sec for $z_{ijk;e}$}\\
\hline
CUDA(GPU)                 & $125.0$ & $116.0$& $91.4$ & $87.2$ & $69.5$ &$77.2$\\
OpenCL(GPU)              & $110.0$ & $83.4$& $84.9$ & $76.6.0$ & $64.6$ &$62.5$\\
OpenMP(16 threads)   & $7.2$ & $8.7$& $7.2$ & $10.4$ & $7.2$ & $6.7$\\
OpenCL(16 threads)   & $6.2$ & $5.8$& $5.1$ & {\color{red}$13.5$} & $4.2$ & $3.7$\\
 \hline
\end{tabular}
}
\end{center}
\caption{Overall timings and performances for different devices and threading systems.\label{tab:perf}}
\end{table}
Thanks to OCCA, the computations were done both on the GPU and on the CPU using common
computational kernels. CPU to GPU speedups of around 10 were observed on all computations:
this is about  the ratio  of memory bandwidths between GPU (200
GB/sec) and the CPU (20 GB/sec). Higher speedups were
obtained at higher orders thanks to the more computationally intensive nature of
higher order computations.

Wall clock times essentially depend on the device: CUDA and OpenCL
give similar performances on the GPU. On the CPU, both OpenMP and OpenCL were run on 16
threads using the hyper-threading capacities of our 8-core processor.

Global performance of the code is similar on the GPU, whether OpenCL or
CUDA is used as threading system. OpenCL is indeed slightly more
efficient, especially for computing the residual.

On the CPU, OpenMP outperforms  OpenCL, except for $n=5$ and $n=7$
where either $r'_{ijk;e}$ ($n=7$) or $z_{ijk;e}$ ($n=5$) shows up
extraordinary performances with OpenCL. At those respective orders,
both the  kernels act on array of size $8 \times 8 \times 8$.
This array size being an exact fraction of a cache line,
memory bandwidth is essentially doubled, reaching about $14$ GB/sec.
Note that padding our arrays to a size that is a power
of two could certainly improve performances at all orders.
OpenCL seems to be better at loop vectorization as well, both on CPU and on GPU.

\section{Conclusions}
In this paper, we have shown that implicit time-steps with over 20 millions
of degrees of freedom in less than half of a second
was possible on an off-the-shelf \$500 GPU.
One of the important features of the method is the two-scale
preconditioner with its two parts that can be computed simultaneously on the
host and the device.

Another important aspect that we have pointed
out, and that significantly differs from the usual way of dealing
with finite element codes, is that it is more interesting to re-compute
quantities like geometrical factors on the fly than to store them in
global memory. This is essentially because GPU computing
is more memory bound  when compared to CPU computing when considering bandwidth
available to each processing element. Therefore, it is
more important to reduce the amount of data that is transferred
from the global memory on a GPU than on a CPU.

Finally, the use of the common kernel language OCCA
allows us to change the device and the thread model in a very simple
fashion i.e. using the same code for all platforms. The OCCA kernel
language is essentially a ``C'' language with extra decorations, which
makes it very readable to computational scientists.

One order of magnitude speedups  is observed between the code
running on the CPU and on the GPU. In both cases, we are able to use
about 10\% of the peak performance of the device, which is to our best
knowledge, close to the best one can get.

This paper is essentially about a fast Poisson solver.
On the one hand the Poisson equation solver is a rather simple proxy application, but on the other hand a Poisson solver is the building block
of more complex models such as incompressible Navier-Stokes equations
or acoustics equations. In further work, non-conforming meshes as well
as incompressible fluids will be considered.

\section*{Acknowledgements}
TW \& RG acknowledge partial support for this research from DOE and ANL (ANL Subcontract
No. 1F-32301 on DOE grant No. DE-AC02-06CH11357), as well as ONR (grant No. N00014-13-1-0873).

\bibliographystyle{plain}
\bibliography{twoscale_tim}

\end{document}